\documentclass[PASP]{emulateapj}

\slugcomment{PASP Accepted}

\shorttitle{The NICMOS Polarimetric Calibration}
\shortauthors{Batcheldor et al.}

\begin{document}


\title{The NICMOS Polarimetric Calibration\altaffilmark{1}}

\author{D. Batcheldor,\altaffilmark{2} A. Robinson,\altaffilmark{2} D. Axon,\altaffilmark{2} D. C. Hines,\altaffilmark{3} 
W. Sparks\altaffilmark{4} \& C. Tadhunter\altaffilmark{5}}

\altaffiltext{1}{Based on observations made with the NASA/ESA Hubble Space Telescope obtained at the 
Space Telescope Science Institute, which is operated by the Association of Universities for Research in Astronomy, Incorporated, under 
NASA contract NAS 5-26555.}
\altaffiltext{2}{Department of Physics, Rochester Institute of Technology, 84 Lomb Memorial Drive, Rochester, NY, 14623, USA.
email: dpbpci@astro.rit.edu}
\altaffiltext{3}{Space Science Institute, 4750 Walnut Street, Suite 205, Boulder, CO 80301, USA}
\altaffiltext{4}{Space Telescope Science Institute, 3700 San Martin Drive, Baltimore, MD, 21218, USA}
\altaffiltext{5}{Department of Physics \& Astronomy, University of Sheffield, Sheffield, S3 7RH, UK}



\begin{abstract}
The value of accurately knowing the absolute calibration of the polarizing elements in the Near Infrared Camera and Multi-Object 
Spectrometer (NICMOS) becomes especially important when conducting studies which require measuring degrees of polarization of close to 
1\% in the near infrared. We present a comprehensive study of all previously observed polarimetric standards using the NIC2 camera on 
NICMOS. Considering both pre- and post-NICMOS Cooling System observations we find variations in the polarimetry consistent with the 
effects of sub-pixel mis-alignments and the point spread function. We also measure non-zero results from unpolarized standards indicating 
an instrumental polarization of $p\approx1.2\%, \theta\approx88\degr$. The lack of polarized and unpolarized standard stars with which to 
perform a comprehensive calibration study means we cannot be confident that the current calibration will be effective for a number of 
recent large NICMOS GO programs. Further observations of polarimetric standards are needed in order to fully characterize the behavior of 
NICMOS at around $p=1\%$.
\end{abstract}




\section{Introduction}

While photometry and spectroscopy provide information on spatial distributions, chemical compositions, and dynamics, polarimetry adds 
a valuable extra data dimension that can be used to determine the geometry of astronomical objects and the properties of interstellar 
particles \citep{aandm85,hranda94,kas03}. Polarized light may intrinsically originate from the emission process or it may be induced via 
interactions with a diffuse medium \citep{coll93,land02}. In either case, the preferred orientation of the electromagnetic vector, as well 
a quantitative measure of that preference, can be determined by placing polarizing elements in the light paths of detectors and deriving 
the Stokes parameters \citep{chan60}.

From the ground some care is needed in determining the Stokes parameters as the atmosphere introduces photometric variations with 
amplitudes similar to those expected from the polarized light. Space based polarimeters are not exposed to these anomalies and 
can use multiple polarizing filters to gather the required information. However, multichannel polarimeters \citep{smandf75}, which are yet 
to be employed on space based polarimeters, do reduce some of the complexities of the data reduction through synchronous recovery of the 
Stokes parameters. In fact, with enough care, ground-based polarimeters can measure fractional polarizations of $10^{-6}$ or less 
\citep{hough05}.

Dust frequently enshrouds many interesting astronomical objects and inhibits the transmission of optical wavelengths. This effect 
declines as the wavelength of the light increases, thus many studies are carried out in the infrared. It is therefore of great advantage 
to be able to carry out polarimetric studies with the Near Infrared Camera and Multi-Object Spectrometer (NICMOS) on board the 
{\it Hubble Space Telescope} ({\it HST}).

\begin{deluxetable*}{lccccccc}
\tablecaption{Details of Observations \label{tab:sample}}
\tablewidth{0pt}
\tablehead{
\colhead{Target} &\colhead{Type} &\colhead{RA (J2000)} & \colhead{Dec (J2000)} & \colhead{Epoch} & 
\colhead{Orientation (\degr)} &\colhead{PID}
}
\startdata
CHA-DC-F7 & B5V (Pol.)     & 10 56 12.91 & -76 35 54.2 & 1997-09-13 &  -36.48 & 7692 \\   
          &                & 10 56 12.03 & -76 35 52.0 & 1998-04-20 & -179.00 & 7958 \\
          &                & 10 56 12.03 & -76 35 52.0 & 1998-07-05 & -106.49 & 7692 \\
          &                & 10 56 12.13 & -76 35 53.2 & 2002-09-02 &  -50.83 & 9644 \\
          &                & 10 56 12.15 & -76 35 53.3 & 2003-05-20 & -160.83 & 9644 \\
G191B2B   & DAw (Unpol.)   & 05 05 30.80 & +52 49 52.9 & 1997-12-24 &  -74.32 & 7904 \\
HD64299   & A1V (Unpol.)   & 07 52 25.61 & -23 17 50.3 & 1997-09-01 &   12.71 & 7692 \\
HD283812  & A2V (Pol.)     & 04 44 25.16 & +25 31 43.1 & 1997-09-28 &   14.52 & 7692 \\
          &                & 04 44 25.16 & +25 31 42.4 & 1997-12-04 &  -47.48 & 7692 \\
HD331891  & A4III (Unpol.) & 20 12  2.29 & +32 47 45.2 & 1997-09-01 &  -93.57 & 7692 \\
          &                & 20 12  2.08 & +32 47 42.2 & 1998-04-26 &   32.55 & 7958 \\
          &                & 20 12  2.11 & +32 47 43.5 & 2002-09-09 & -109.63 & 9644 \\
          &                & 20 12  2.11 & +32 47 43.5 & 2003-06-08 &    0.57 & 9644 \\ \hline
\enddata
\tablecomments{Types taken from SIMBAD (http://simbad.u-strasbg.fr/Simbad), all other data are as they appear in the {\it HST} archive 
(http://archive.stsci.edu/hst/). (Pol.) and (Unpol.) after the  stellar type refers to polarized and unpolarized standards respectively.
}
\end{deluxetable*}

As with many instruments, NICMOS executes programs that cover a wide range of required accuracies and sensitivities. It is therefore 
essential to probe the polarimetric capabilities of NICMOS, especially in low polarization targets, and assess any signatures that may be 
introduced by the instrument itself. We present here such an assessment based on the previous efforts of, for example, \citet{hsands00} 
and \citet{hines02}. By examining the on-line data archive for NICMOS we can directly compare the results from polarimetric standards 
whose degree and orientation of polarization have been determined during other investigations. The study is limited to include data 
obtained with the NIC2 ($\Delta\lambda=1.9-2.1\micron$) camera as the number of previous NIC1 ($\Delta\lambda=0.8-1.3\micron$) studies, 
excluding the calibration data, is low. In addition, these NIC1 programs generally probe polarizations of $\sim40\%$, i.e., \#7264, where 
high accuracies are not necessarily needed. However, future works similar to those carried out here may also be prudent for NIC1. 

In \S~\ref{obs+dr} we describe the data and its subsequent reduction. The methods used in determining the degree and orientation 
of polarization are outlined in \S~\ref{ptheta}. The results are presented in \S~\ref{results} before being discussed and concluded in 
\S~\ref{discussion} and \S~\ref{cons} respectively.

\section{Observations and Data Reduction}\label{obs+dr}

During the {\it HST} second servicing mission in 1997, the Faint Object Spectrograph was replaced by NICMOS. This instrument
operates from $0.8-2.5\micron$ using three independent cameras. The setup allows for broad, medium and narrow band imaging, coronographic 
imaging, broad-band imaging polarimetry, and grism spectroscopy. NIC2, the camera focused on in this study, is a $256\times256$ HgCdTe 
array with a 0\farcs075 pixel scale giving a 19\farcs2$\times$19\farcs3 field of view \citep{thom98}. After installation on {\it HST} the 
dewar was allowed to be brought up to operating temperature. Unfortunately, this temperature was not tested on the ground and the 
subsequent ice expansion provided enough dewar deformation to allow contact between one of the optical baffles and the vapor cooled 
shield. The resulting heat sink not only altered the foci of the three cameras, but also depleted the cryogen by the beginning of 1999 
(two and a half years ahead of schedule). However, a mechanical cryocooler, the NICMOS Cooling System (NCS), which was successfully 
installed in March 2002, has since brought NICMOS back to life and restored the infrared capabilities of {\it HST}. While the instrument 
performance is comparable pre- and post-NCS \citep{hines02}, it must be noted that NICMOS now displays a different set of characteristics. 
The reason for these changes are at present unclear, however, for individual observations, these effects can be incorporated as
changes in the transmission efficiencies of polarizing elements. One must also be aware that different quadrants of the array may exhibit 
different bias levels introduced by resetting the array, plus additional bias offsets that are related to operating temperature and main 
bus voltage \citep{berg05}. While this ``pedestal effect'' has been reduced by modified flight software, it has not been entirely 
eliminated and may lead to inconsistent polarization measurements. More specific details can be obtained from the NICMOS Instrument 
Handbook \citep{sch05}.

A comprehensive search of the NICMOS data archive has been performed with the aim of analyzing all observed polarimetric standard stars. 
There are also a number of extended sources (e.g. CRL 2688 - the Egg Nebula) available for use in a calibration study. However, the 
ground truth for such objects is useless as the measured polarizations are dependent on resolution. They also generally exhibit 
a high degree of polarization and are often variable. Consequently, such objects will not be conducive to determining the polarimetric 
accuracies for the low polarization targets ($\lesssim5\%$) with which we are concerned. A prerequisite for the standards to be included 
in the sample is independent polarization measures from well calibrated and characterized polarimeters. The polarimetric standards found, 
and associated details, are presented in Table~\ref{tab:sample}. In addition to the data in Table~\ref{tab:sample} there are also pre-NCS 
data for two unpolarized standards, HD10700 and HD30652 \citep{landb89,leroy93}. Unfortunately, the program for which these observation 
were made (\#7614) called for the investigation of circumstellar structures. The actual stars are therefore hopelessly overexposed for 
accurate study here.

The data and calibration files were retrieved and passed through the latest versions of the {\it calnica} and {\it calnicb} pipelines 
where appropriate. Each non-destructive readout of individual observations was examined. The total counts in each {\tt multiaccum} frame 
were inspected as a function of the total exposure time. Deviations of this curve of growth from linear, especially at very early times, 
were deemed to be evidence for the existence of signal persistence. In the cases where the telescope was offset between exposures, the 
areas of the array exposed to source photons in the previous visit were also examined for persistence. The individual exposures were also 
checked for any remaining cosmic rays or hot pixels not flagged by the reduction routine.

The {\it apphot} package in IRAF\footnote{IRAF is distributed by the National Optical Astronomy Observatories, which are operated 
by the Association of Universities for Research in Astronomy, Inc., under co-operative agreement with the National Science foundation.} 
was used in order to perform aperture photometry on each polarized standard. Circular apertures with radii from 0.5 to 50.5 pixels (in 1 
pixel intervals) were centroided on each standard. The observable point spread function (PSF) was comfortably included in the outer 
aperture. In each frame the sky was sampled (and removed from the target photometry) using a disk whose inner annulus was defined by the 
edge of the observable PSF, and whose outer annulus was defined by the edge of the array. In every case the sky aperture was large enough 
to avoid the irregularities introduced by the small number statistics associated with Poisson noise. 

\section{Determining P and Theta}\label{ptheta}

The methodologies behind extracting the degree and orientation of polarization have been specifically addressed for NICMOS by many 
authors \citep{msanda98,mandh99,sanda99,hsands00,hines02}. However, we briefly revisit the linear technique for the case of three 
non-ideal polarizers here. 

The reduced, sky subtracted, instrumental counts (in ${\rm~DN~s^{-1}}$ space) through each polarizing element are used to define an 
observed intensity vector of the form $a=[I_1,I_2,I_3]$. The degree and orientation of polarization ($p$ and $\theta$) are defined by the 
Stokes parameters which also define a vector of the form $b=(I,Q,U)$. The two vectors $a$ and $b$ are simply related to each other by the 
linear expression $[C]b=a$, where $[C]$ is a matrix describing the characteristics of the $k^{\rm th}$ polarizer, namely the polarizer 
efficiency ($\eta_k$), the actual orientation (in radians) of the polarizer ($\phi_k$), the fraction of the light transmitted in the 
parallel direction ($t_k$), and the fraction of light transmitted in the perpendicular direction ($l_k$). The transmission coefficients 
are then described by;

\begin{equation}\label{equ:transcoeffs}
X_k = \frac{1}{2}t_k(1+l_k)\end{equation}
\begin{equation}Y_k = X_k(\cos{2\phi_k})\end{equation}
\begin{equation}Z_k = X_k(\sin{2\phi_k})\end{equation}

allowing the matrix $[C]$ to be defined as follows:

\begin{equation}
C =  \left[
  \begin{array}{ c c c}
     X_1 & \eta_{1}Y_1 & \eta_{1}Z_1 \\
     X_2 & \eta_{2}Y_2 & \eta_{2}Z_2 \\
     X_3 & \eta_{3}Y_3 & \eta_{3}Y_3 \\
  \end{array} \right]	
\end{equation}

The linear expression $[C]b=a$ can then be solved for $I, Q$ and $U$ using, for example, LU decomposition. We can then solve for $p$ and 
$\theta$ using Equations~\ref{equ:p} and \ref{equ:theta}.

\begin{equation}\label{equ:p}
p = 100\% \times \frac{\sqrt{Q^2+U^2}}{I}
\end{equation}

\begin{equation}\label{equ:theta}
\theta = \frac{1}{2}\arctan{\frac{U}{Q}}
\end{equation}

In Equation~\ref{equ:theta} a 360\degr arctangent function is assumed. In addition, the orientation of the frame has to be subtracted 
from $\theta$ in order to retrieve the on sky position angle.

We have coded the above process in IDL in order to solve for $p$ and $\theta$ given three lists of apertures and photometric data. This 
process has also been coded by \citet{mandh99} into the IDL routine 
{\it polarize.pro}.\footnote{http://www.stsci.edu/hst/nicmos/tools/polarize\_tools.html} Their routine takes the three polarized images 
and produces two dimensional maps of $p$ and $\theta$, whereas the code used here produces the measured polarization through defined 
apertures. The two methods are precisely consistent when tested on flat (polarized) data, i.e., one value of $I$ for each 
polarizer across the entire array. 

\begin{deluxetable}{lccccc}
\tablecolumns{6}
\tablecaption{Polarimetric Calibration Parameters\label{tab:coeffs}}
\tablewidth{0pt}
\tablehead{
                 &                      &                    &                 & \multicolumn{2}{c}{$t_k$}            \\
                 &                      &                    &                 & \multicolumn{2}{c}{\line(1,0){60}}   \\
\colhead{Filter} & \colhead{$\theta_k (\degr)$} & \colhead{$\eta_k$} & \colhead{$l_k$} & \colhead{-NCS} & \colhead{+NCS}
}
\startdata
POL0L   &   8.84 & 0.7313 & 0.1552 & 0.8981 & 0.8779 \\
POL120L & 131.42 & 0.6288 & 0.2279 & 0.8551 & 0.8379 \\
POL240L & 248.18 & 0.8738 & 0.0673 & 0.9667 & 0.9667 \\
\enddata
\tablecomments{``-'' and ``+'' refer to pre- and post-NCS co-efficients respectively.
}
\end{deluxetable}

In addition to the aforementioned process, we have also included an estimation of the errors associated with $p$ and $\theta$ 
introduced by the variance in the count rate. The estimation is taken from \citet{sanda99} and is defined by the variance and covariance 
in the ($Q,U$)-plane.

\subsection{The Polarimetric Calibration Parameters}

The first estimates of the polarimetric calibration parameters were presented by \citet{hsandl97}. Preflight thermal vacuum tests of the 
polarizing optics found each element to have a unique polarizing efficiency and position angle offset. The values of $\phi_k$, $\eta_k$, 
$l_k$ and $t_k$ were estimated by attaching a polarizer to the Calibrated InfraRed sourCE (CIRCE). This provided uniform illumination 
across the entire array at a known position angle and magnitude. Following the on orbit observations of the polarized standard CHA-DC-F7 
and the unpolarized standard HD331891, values for $t_k$ were adjusted based on exposures from two separate roll angles and from two 
separate epochs \citep{hines98,hsands00}. Using the pre-NCS coefficients from Table~\ref{tab:coeffs}, values of $p$ for CHA-DC-F7 were 
found to be within $\sim0.2\%$ of the ground based values. Post-NCS the values of $t_k$ were again adjusted (0.8774, 0.8381, 09667; Hines 
2002) and used in such studies as \citet{umandm05}. The coefficient matrix used in this study has been constructed from the parameters in 
Table~\ref{tab:coeffs} using the latest revised values of $t_k$ \citep{hines05}.

\begin{figure*}[t]
\plotone{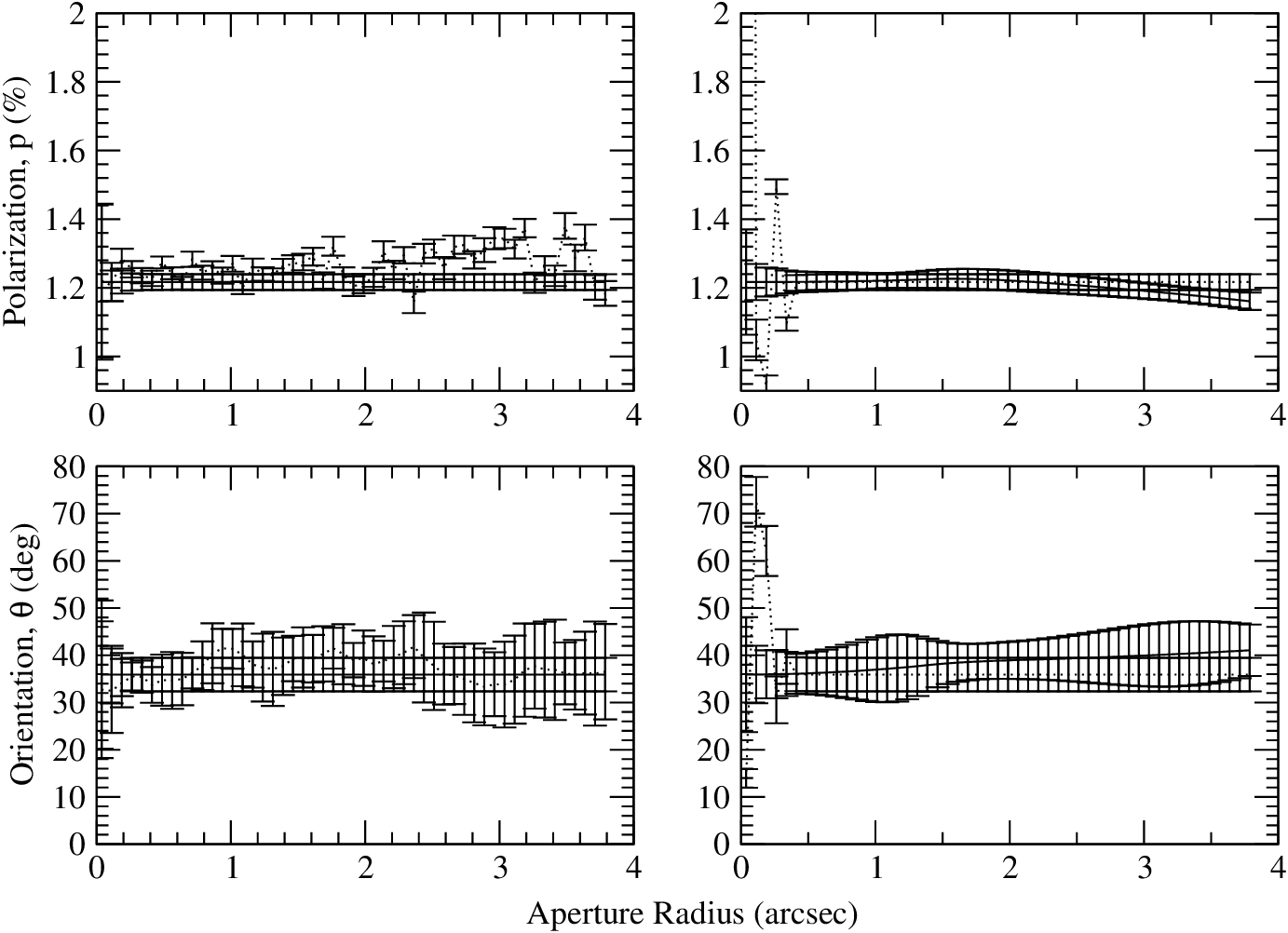}
\caption[Testing the Routine]{
Results from testing the polarimetry routine. ({\it Left column}) From a Gaussian surface brightness profile with (dotted line) and 
without (solid line) Poisson noise. ({\it Right column}) Using the PSFs as generated by {\tt TinyTim}. The solid lines are for perfectly 
aligned frames whereas the dotted lines show the effects of sub-pixel mis-alignment using a Gaussian profile. 
}\label{fig:tests}
\end{figure*}

\subsection{Testing the Routine}

The IDL routine was extensively tested in order to check its performance against known (simulated) polarizations. The testing also 
provided examples of how the instrument itself may affect the observed polarization (e.g., noise and the PSF). We have generated data, 
mapped onto the NICMOS pixel grid, which simulates a star with a Gaussian surface brightness profile (both with and without the effects of 
Poisson noise), a point source with a {\tt TinyTim} PSF \citep{kandh04}, and the effects of sub-pixel shifts between the pointing of the 
three polarizers. Small spatial shifts between polarizing elements are not unexpected and may allow flux gathered in one frame to fall 
into an inter-pixel gap in another.

The resulting polarization profiles (which all use the post-NCS transmission coefficients for NIC2), i.e., the polarization as measured 
through progressively larger apertures, are presented in Figure~\ref{fig:tests}. It can be seen that the routine behaves as one would 
predict. Generally there are variations plus larger errors on the small scale, and constant results for large apertures. The addition of 
the {\tt TinyTim} PSF is seen to increase the small scale variations and errors. Introducing sub-pixel shifts allows the small scale 
variations to grow. In all noiseless cases the profile behavior is stable in apertures larger than 0\farcs5. The best polarization 
estimates for real data are therefore likely to come from large apertures enclosing most of the flux.

In all cases, to get appropriate estimates for the effects of Poisson noise, we have replicated the signal to noise ratios typically 
seen in the standards ($\sim50$) as well as the expected polarizations ($\sim1.2\%$).

\section{Results}\label{results}

We now pay close attention to each of the targets listed in Table~\ref{tab:sample}. The results and details of previous ground based  
polarimetry observations, as well as the archived NICMOS results, are presented individually, summarized in Table~\ref{tab:results} 
and compared in Figure~\ref{tab:results}. 

In order to correct for the wavelength dependence of polarization \citep{smandf75} and allow a comparison between these NICMOS 
measurements and ground based results, one can use the ``Serkowski curve'' \citep{serk73,wil82} to find $p(2.05)$. This notation 
corresponds to the value of $p$ (\%) at the $2.05\micron$ wavelength of NIC2. However, in both cases where this may be necessary, there 
are direct measurements at $p(2.04)$ by \citet{whit92}.

\begin{deluxetable*}{lcccc|cccc}
\tablecolumns{8}
\tablecaption{Results of This and Previous Studies \label{tab:results}}
\tablewidth{0pt}
\tablehead{
                 & \multicolumn{4}{c}{Previous}                                 & \multicolumn{4}{c}{This Study}     \\
                 & \multicolumn{4}{c}{\line(1,0){195}}                           & \multicolumn{4}{c}{\line(1,0){195}} \\
\colhead{Target} & \colhead{$p_{pre} (\%)$} & \colhead{$\theta_{pre} (\degr)$} & \colhead{$p(\lambda_{max})(\%)$} & \colhead{Ref.} & 
\colhead{Epoch} & \colhead{R('')} & \colhead{$p (\%)$} & \colhead{$\theta (\degr)$} 
}
\startdata
CHA-DC-F7 & $1.19\pm0.01$ & $126\pm4$ & 5.98 & 1   & 1997-09-13 & 0.34 & $0.94\pm0.03$ & $116\pm7$ \\   
          &               &           &      &     & 1998-04-20 & 0.34 & $2.22\pm0.03$ &  $86\pm4$ \\
          &               &           &      &     & 1998-07-05 & 0.34 & $0.95\pm0.03$ & $116\pm2$ \\
          &               &           &      &     & 2002-09-02 & 0.34 & $1.05\pm0.02$ & $119\pm5$ \\
          &               &           &      &     & 2003-05-20 & 0.34 & $1.19\pm0.03$ & $111\pm4$ \\
G191B2B   & $0.09\pm0.05$ & $157\pm1$ & n/a  & 2,3 & 1997-12-24 & 0.26 &  $1.2\pm0.4$  & $90\pm5$  \\
HD64299   & $0.15\pm0.03$ & \nodata   & n/a  & 2   & 1997-09-01 & 0.34 & $1.46\pm0.06$ & $67\pm4$  \\
HD283812  & $1.31\pm0.07$ &  $35\pm2$ & 6.29 & 1   & 1997-09-28 & 0.79 & $3.11\pm0.08$ & $121\pm3$ \\
          &               &           &      &     & 1997-12-04 & 0.79 & $1.40\pm0.03$ & $35\pm4$  \\
HD331891  & $0.04\pm0.03$ & $79\pm1$  & n/a  & 2,3 & 1997-09-01 & 0.34 & $1.32\pm0.02$ & $106\pm4$ \\
          &               &           &      &     & 1998-04-26 & 0.34 & $1.19\pm0.02$ & $97\pm5$  \\
          &               &           &      &     & 2002-09-09 & 0.34 & $1.41\pm0.02$ & $90\pm3$  \\
          &               &           &      &     & 2003-06-08 & 0.34 & $0.77\pm0.02$ & $79\pm3$  \\ \hline
\enddata
\tablecomments{
References: 1. \citet{whit92}. 2. \citet{turn90}. 3. \citet{seandl92}. The values of $p_{pre}$ and $\theta_{pre}$ 
are those taken from the literature {\it after} the $p(\lambda_{max})$ correction. R('') refers to the aperture radius through which the 
observation was made. 
}
\end{deluxetable*}

We have considered photometry from data which have also had the model PSFs, as generated by {\tt TinyTim}, removed using the 
Lucy-Richardson deconvolution. In the test cases (CHA-DC-F7 and HD331891) a change of less than 0.1\% in $p$ was seen inside the PSF. 
Using large apertures there was no difference as all of the flux was enclosed without the need for deconvolution. However, the advantage 
of de-convolving the PSF comes when attempting to visually identify remaining hot pixels and cosmic rays. The extra flux that may be 
introduced by such pixels hiding in the artifacts of the PSF will have a detrimental effect on the photometry and therefore the 
measured polarization. The residuals between the PSF deconvolved frame and frame containing the PSF were therefore carefully examined 
for further evidence of hot pixels and cosmic rays. Two dimensional interpolation is used to replace any such features. All of the 
remaining polarimetry was carried out with PSF deconvolution.

In this study, where we can see the polarimetric behavior with radius, it is not obvious which aperture will be consistent with 
previous studies; photometric aperture sizes have not been published. Here we choose to inspect each frame in order to report the 
measured polarization in an aperture that includes all of the observed flux distributed across the array by the PSF. The aperture size is 
also reported in Table~\ref{tab:results}. 

\subsection{CHA-DC-F7}\label{chadcf7}

The polarimetric properties of CHA-DC-F7 were originally reported by \citet{whit92} before being used in the previous NICMOS polarimetry 
studies of \citet{hsandl97}, \citet{hines98}, \citet{hsands00} and \citet{hines02}. \citet{whit92} report, from ground based 
measurements, $p(\lambda_{max})=5.98\%$ at $0.55\micron$ with an instrumental polarization of $0.03\%$, and $p(2.04)=1.19\%\pm0.01\%$ at 
$\theta=126\degr\pm4\degr$ which we use in this comparison. The NICMOS polarimetry studies \citep{hsands00} report 
$p(2.05)=0.97\%\pm0.2\%$ at $\theta=119\degr\pm6\degr$ and $p(2.05)=1.00\%\pm0.2\%$ at $\theta=119\degr\pm6\degr$ for the 1997 and 1998 
epochs respectively. 

Exposures of 13.95 seconds were made, at every epoch, through each of the three polarizing elements. With the exception of the 1997 and 
July 1998 epochs all observations used a four point dither pattern, each frame being offset by 2\farcs3. All data in this study are 
consistent with previous polarization estimates apart from the April 1998 epoch. We note that of all the data for CHA-DC-F7 these 
observations are closest ($\sim50$ minutes) to a South Atlantic Anomaly passage and are the only pre-NCS observation to be dithered. While 
the number of cosmic ray hits does not appear to be higher than in the other epochs, we have noticed large differences in the photometry 
from different parts of the array. Figure~\ref{fig:dither} demonstrates the difference between photometry obtained from the individual 
non-destructive readouts from the dithered and non-dithered data. It can be seen that there is a larger spread in the photometry from the 
dithered data, which in turn could lead to a greater amount of polarization due to the greater flux differences between the polarizers.
This effect is not present in post-NCS observations and suggests that the pedestal effect is non-negligible in pre-NCS dithered 
polarimetry. It is well known that the NCS produces much more stable temperatures for the array \citep{srands03,arr05}, which is why the 
pedestal effects are much less dramatic than when NICMOS was cooled with nitrogen ice. We also note that there are late variations from 
linear ($\sim10$ seconds onward) in the photometric curves of growth. This may be indicative of saturation, or at least non-linearity, but 
we do not find any evidence for this in the Data Quality (DQ) data extensions. DQ values of 3072 are reported which correspond to 
``Pixel containing source'' (1024) and ``Pixel has signal in the 0th read'' (2048), not ``Saturated pixel'' (64) or ``Poor or uncertain 
Linearity correction'' (2)\footnote{See Chapter 2 of the NICMOS Data Handbook, \citet{mandr04}.}. Following on from this we also do not 
see any non-linearity early in the curves of growth. This indicates that persistence is insignificant.

\subsection{G191B2B}\label{g191b2b}

\citet{seandl92} report G191B2B to be an unpolarized standard with $p(0.55)=0.09\%\pm0.05\%$ and $\theta=157\degr$ from observations 
using the ``Two-Holer'' polarimeter. The instrumental polarization is estimated to be 0.05-0.10\%. 

Each NICMOS observation has an exposure time of 23.97s and has been dithered using a three point pattern. Compared to the CHA-DC-F7 
observations the point spacings for these dithers are small. In addition, the spread in the photometric curves of growth is less than 
those seen in Figure~\ref{fig:dither}. Nevertheless the results from this study do show polarization at a level of 1.2\% where 
something much closer to zero is expected.

\subsection{HD64299}\label{hd64299}

HD64299 is listed as an unpolarized standard ($p(\sim~0.55)=0.15\%\pm0.03\%$) by \citet{turn90}. \citet{aandw99} also use this standard in 
order to characterize the instrumental polarization of the ADONIS polarimeter ($p_{ins}(\sim~2.0)\approx1.5\%$).

No dithering was used during each of the 11.96s exposures and there is no evidence for persistence or saturation. However, as with the 
case of G191B2B, we measure a polarization that is inconsistent with zero.

\begin{figure}[t]
\vspace{0.6cm}
\plotone{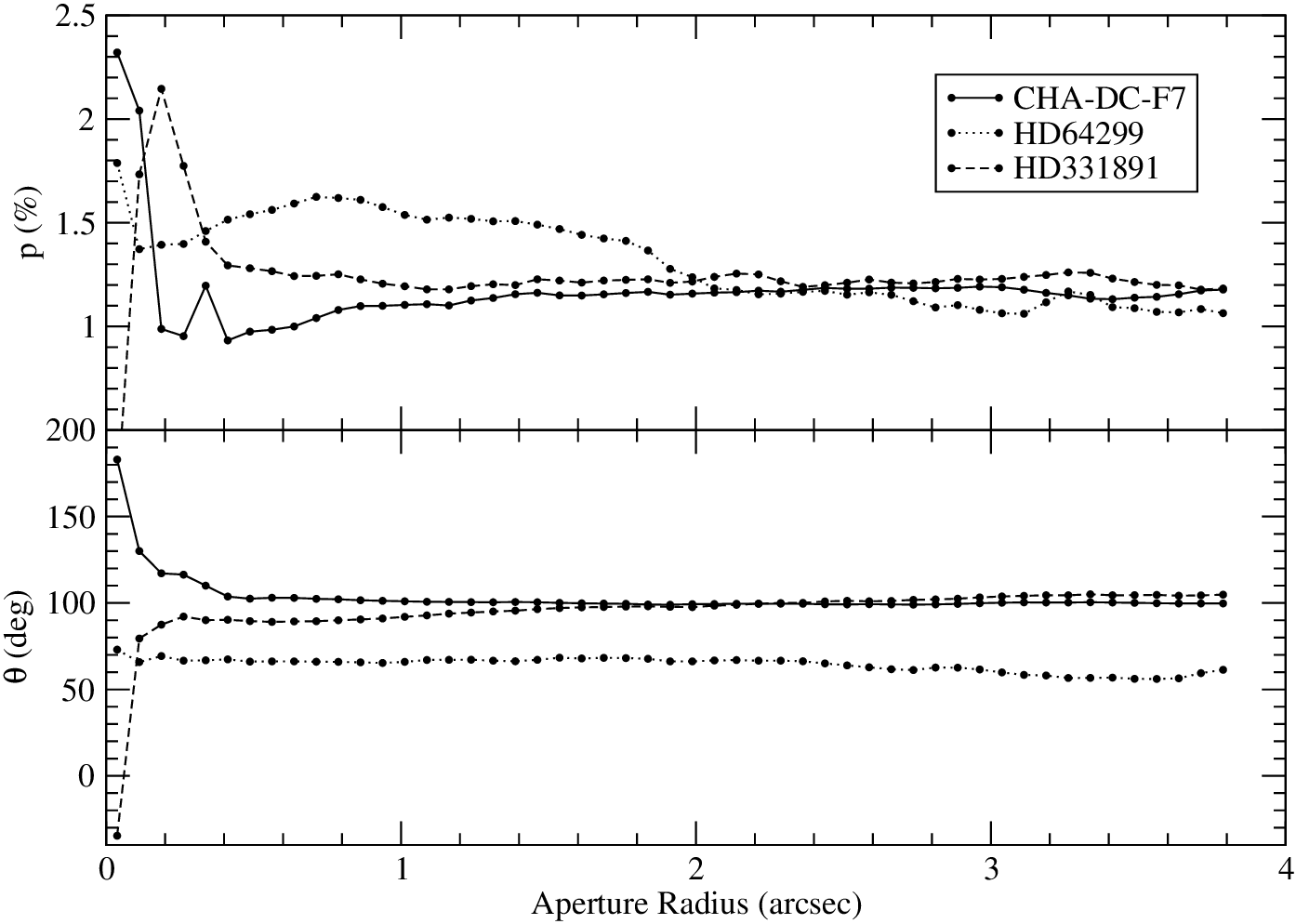}
\caption[Example Polarimetric Curves of Growth]{
Example polarimetric curves of growth. Error bars have been omitted for clarity but are approximately $\pm0.03\%$ in $p$ and $\pm4\degr$ 
in $\theta$.
}\label{fig:rescomp}
\end{figure}

\subsection{HD283812}\label{hd283812}

The polarized properties of HD283812 have been reported to be $p(\sim~0.55)=4\%\pm1\%, \theta=33.8\degr$ and $p(0.55)=6.29\%\pm0.05\%, 
\theta=32\degr\pm1\degr$ by \citet{turn90} and \citet{whit92} respectively. \citet{whit92} also measures $p(2.04)=1.31\%\pm0.07\%$ at 
$\theta=35\degr\pm2\degr$. As HD283812 is an extended source, we use the brightest northern component to begin the centroid for photometry.

\begin{figure}[t]
\vspace{0.6cm}
\plotone{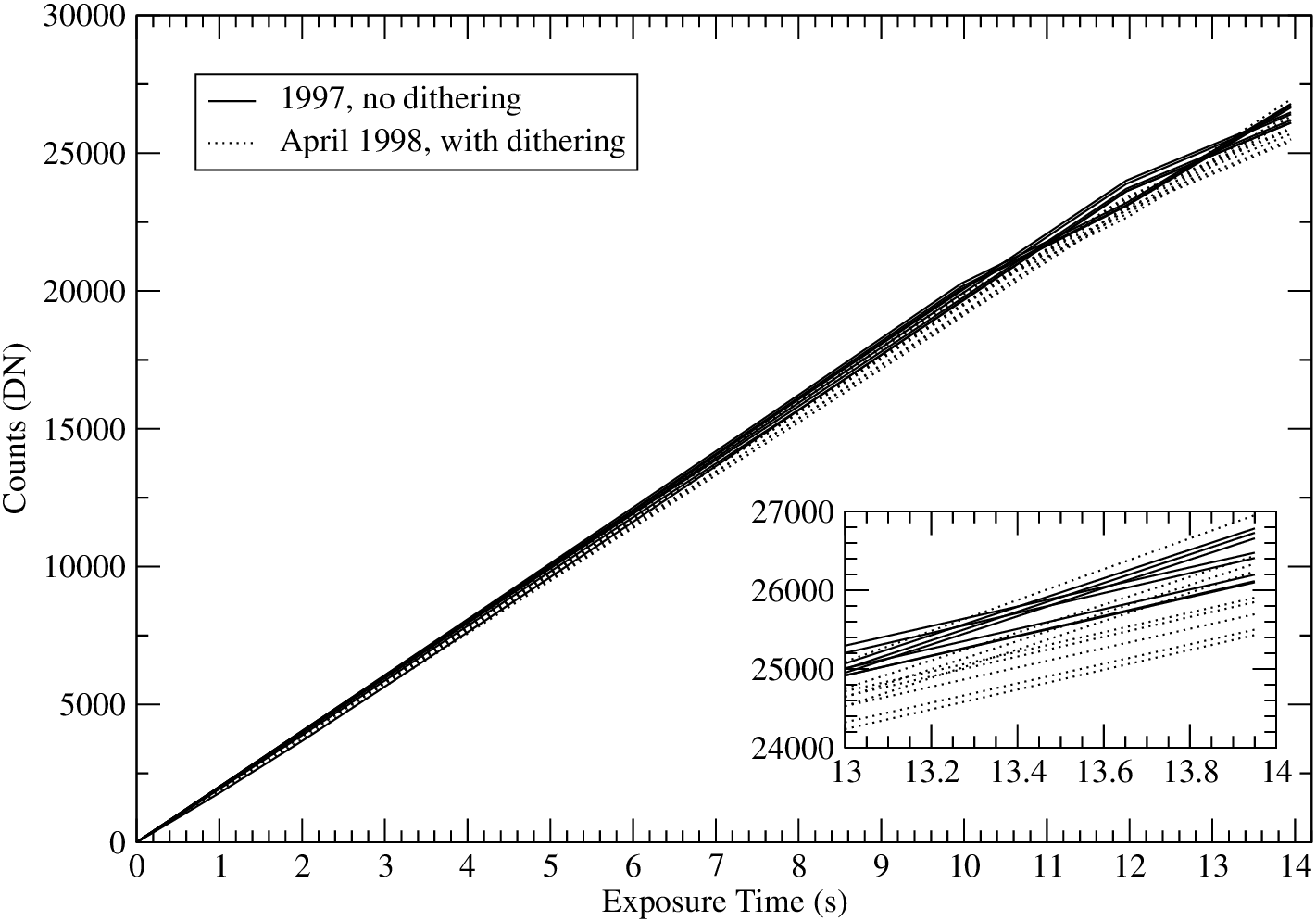}
\caption[Pre-NCS Dithering]{
Comparing non-destructive readouts from dithered and un-dithered data for CHA-DC-F7. Inset: An expanded view from the end of the 
exposures highlighting the photometric differences.
}\label{fig:dither}
\end{figure}

Both epochs of the data for HD283812 had exposure times of 5.98s. Dithering was only applied to the September 1997 epoch data however, 
where the point spacing was large (5\farcs0). The dithered photometric curves of growth show a spread comparable to the dithered data in 
Figure~\ref{fig:dither}, there is no evidence of persistence. We find that the dithered data shows a larger degree and orientation of 
polarization than expected but the non-dithered data is entirely consistent. 

\subsection{HD331891}\label{331891}

\citet{turn90} and \citet{seandl92} report HD331891 (BD32+3739) to be an unpolarized standard with $p(\sim~0.43)=0.04\%\pm0.03\%$. This 
standard was also used in determining the pre- and post-NCS coefficients.

Whereas all exposures were 5.98s, all but the 1997 data have been dithered with a 2\farcs3 four point spacing. There is no evidence of 
persistence and none of the data shows an exaggerated spread in the photometric curves of growth. Nevertheless, all epochs show non-zero 
degrees and orientations of polarization. These results are consistent with the findings of \citet{umandm05} who use the same data with 
the same co-efficients to find $p(2.05)=1.4\%$. They attribute the $\gtrsim1\%$ polarization, i.e., greater than the $\lesssim1\%$ 
instrumental polarization reported by \citet{hsands00}, to ``systematics in the data reduction procedure.'' However, the estimate of 
$p_{ins}\lesssim1\%$ was based upon ground based-thermal vacuum tests and not from on-orbit data.

\section{Discussions}\label{discussion}

As demonstrated by Figure~\ref{fig:rescomp} we can see that the simulated (Figure~\ref{fig:tests}) and real data compare well. In all 
cases the large-scale values of $p$ and $\theta$ are clearly stable. The inner variations due to the residual PSF and sub-pixel 
mis-alignment are also seen, but are insignificant outside of a radius which approximately corresponds to the first airy ring of the PSF 
($\sim0\farcs3$), i.e., as long as the photometric apertures contain all of the flux spread by the PSF, consistent values of $p$ and 
$\theta$ will be measured.

It is clear that there are significant differences in polarizations measured at separate epochs for CHA-DC-F7 and HD283812. It is unlikely 
that these variations are intrinsic to the objects. In both cases it is noted that the data are pre-NCS and dithered. It is also unclear 
why such an anomaly is not seen in the pre-NCS dithered data for HD331891. However, as suggested in \S\ref{chadcf7}, this may be an 
artifact from the pedestal effect. There do exist several IRAF software packages to remove the pedestal effect ({\it pedsky}, 
{\it pedsub}), but as the dither allows a check of the NICMOS polarization properties in each quadrant, we can check to see if this is 
the reason for the discrepant results by performing our routine only on the data from a single pointing. Accordingly we have re-reduced 
the raw data files through the {\it calnica} pipeline and determined the polarization results from each quadrant. The results are 
presented in Table~\ref{tab:pedres}. It can be seen that the individual pointings give polarization measures consistent with un-dithered 
data, and that the pre-NCS pedestal effect can induce variations of $\Delta{p}\approx0.2\%$ and $\Delta\theta\approx15\degr$ between 
different quadrants. Post-NCS the pedestal effect does not appear to have affected the results, however, it is clear that the pedestal 
effect does produce slightly different polarizations in each quadrant of the array.

We have measured polarization in all of the targets which have previously been listed as unpolarized. Such findings suggest 
that there may be a low level of instrumental polarization. From Table~\ref{tab:results} it can be seen that, on average, this 
instrumental polarization has a magnitude $p\approx1.2\%$ at a position angle of $\theta\approx88\degr$. We now demonstrate how 
such an instrumental polarization may be corrected for.

The fact that polarized standards have been measured to be consistent with previous studies (CHA-DC-F7, HD283812) allows us to be 
confident that the results for the unpolarized standards are real, and that we may be seeing a residual instrumental polarization. The 
low number of observed unpolarized standards and the pedestal effect make it difficult to be confident in the exact nature of the 
instrumental polarization, but we can nevertheless make an attempt to correct the data for this. As the degree and orientation of 
polarization are derived in Stokes ($I,Q,U$) space this is also where we can perform an instrumental correction. It is essential that the 
correction is carried out in the ($Q,U$)-plane for each standard and not in the celestial coordinate system. The unpolarized standard data 
from pre- and post-NCS observations are considered separately. While the ($Q,U$) quadrants are consistent between the polarized standards, 
the actual values of $Q$ and $U$ are dependent on $I$. The corrections are therefore averaged (weighted by the errors in $Q$ and $U$) 
across all observed standards after being normalized by the intensity, rather than being simply carried out in the ($Q,U$)-plane. The 
results from the correction are presented in Table~\ref{tab:correct}. It can be seen that the dispersion around the expected and 
previously reported results has been reduced. The $\chi^2$ for the polarized standards about the expected values is now $\sim0.01$ as 
compared to $\sim0.1$ without the ($Q,U$) correction. 

\begin{deluxetable}{ccc|cc}
\tablecolumns{5}
\tablecaption{Pre-NCS Pedestal Check\label{tab:pedres}}
\tablewidth{0pt}
\tablehead{
                   &\multicolumn{2}{c}{CHA-DC-F7}      &\multicolumn{2}{c}{HD283812}       \\
                   &\multicolumn{2}{c}{\line(1,0){75}} &\multicolumn{2}{c}{\line(1,0){75}} \\
\colhead{Quadrant} &\colhead{$p (\%)$} & \colhead{$\theta (\degr)$} &\colhead{$p (\%)$} & \colhead{$\theta (\degr)$} \\
}
\startdata
1 & $1.25\pm0.03$ & $115\pm6$ & $1.49\pm0.03$ & $31\pm2$ \\
2 & $1.12\pm0.03$ & $128\pm7$ & $1.41\pm0.01$ & $24\pm4$ \\
3 & $1.13\pm0.03$ & $113\pm7$ & $1.53\pm0.03$ & $35\pm4$ \\
4 & $1.21\pm0.02$ & $114\pm6$ & $1.31\pm0.03$ & $36\pm2$ \\
\enddata
\end{deluxetable}

\begin{figure}[t]
\vspace{1.0cm}
\plotone{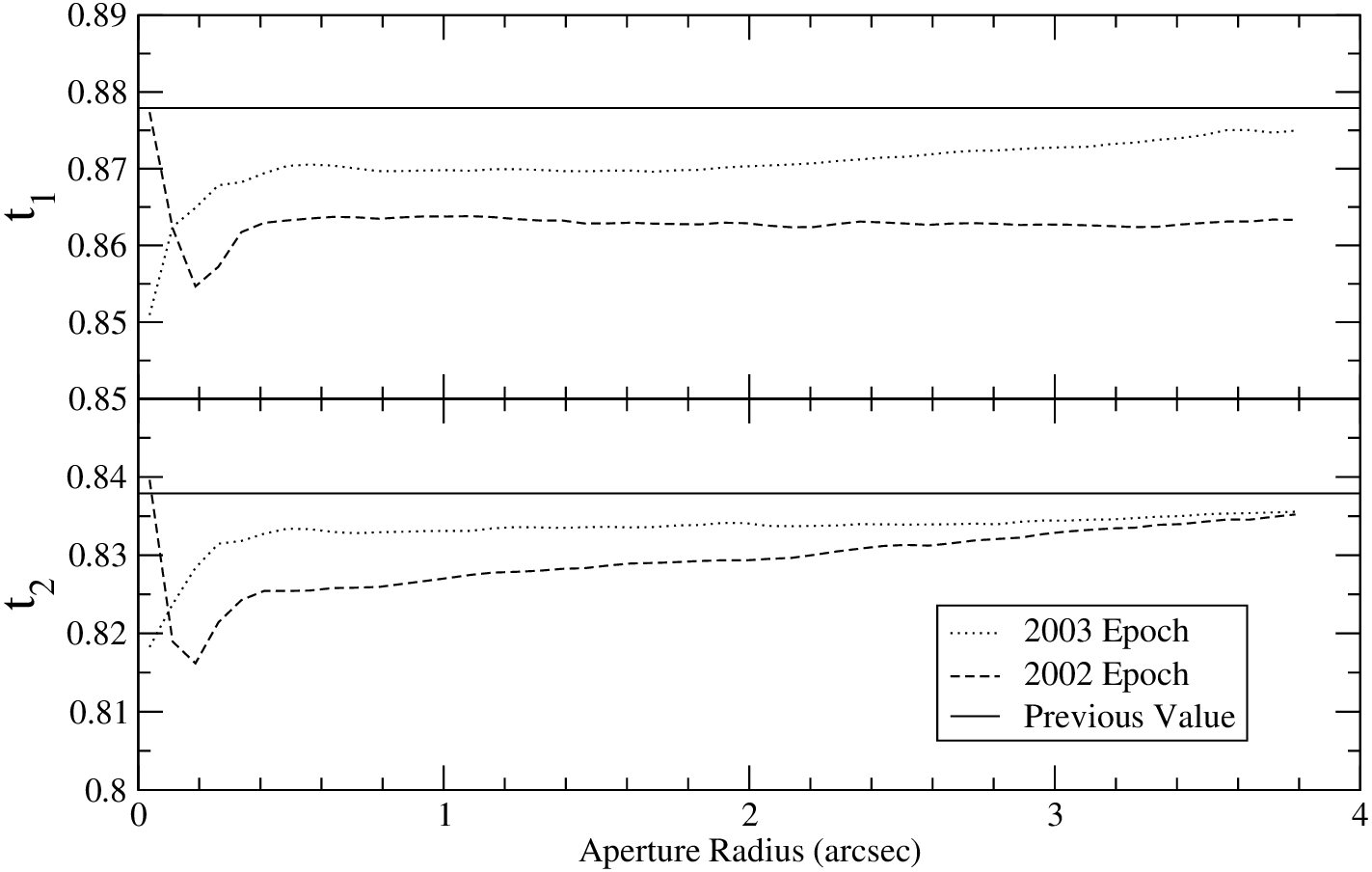}
\caption[The Parallel Transmission Co-efficients]{
The values of $t_k$ required to null the unpolarized standard HD331891. The value of $t_3$ is held constant at 0.9667.
}\label{fig:coeffs}
\end{figure}

It is also possible to null the polarization in unpolarized targets by adjusting the on orbit derived transmission co-efficients.
As Q = U = 0 in unpolarized sources, $t_k$ can be derived by considering $I_k=IX_k$ and fixing $t_3 = 0.9667$. Solving $t_1$ and $t_2$ 
for $p=0\%$ and $\theta=0\degr$ in HD331891 across all photometric apertures results in the co-efficient profiles presented in 
Figure~\ref{fig:coeffs}. It can be seen in Figure~\ref{fig:coeffs} that in the central most aperture the co-efficients derived from the 
2002 epoch (dashed line) are consistent with the co-efficients quoted in Table~\ref{tab:coeffs}. Similar small scale variations to that 
of the observed polarimetric curves of growth are also seen. Taking an average of the derived 2003 epoch co-efficients, outside of the 
inner arc-second, and carrying through the photometric errors, gives us $t_1 = 0.8717\pm0.0005$ and $t_2 = 0.8341\pm0.0005$. 
Applying these adjusted co-efficients (which are within 1\% of the original co-efficients) to the 2003 data for CHA-DC-F7 
produces a typical change of $\delta{p}\approx0.1\%$ and $\delta{\theta}\approx3\degr$ over the values of $p$ and $\theta$ derived from 
the original co-efficients.

\begin{deluxetable}{lccc}
\tablecolumns{4}
\tablecaption{Correcting for Instrumental Polarization \label{tab:correct}}
\tablewidth{0pt}
\tablehead{
\colhead{Target} & \colhead{Epoch} & \colhead{$p_{corr} (\%)$} & \colhead{$\theta_{corr} (\degr)$} 
}
\startdata
CHA-DC-F7 & 1997-09-13 & $1.24\pm0.03$  & $128\pm7$  \\   
          & 1998-04-20 & $1.25\pm0.03$* & $117\pm5$  \\
          & 1998-07-05 & $1.21\pm0.04$  & $133\pm5$  \\
          & 2002-09-02 & $1.21\pm0.03$  & $109\pm7$  \\
          & 2003-05-20 & $1.12\pm0.04$  & $124\pm6$  \\
G191B2B   & 1997-12-24 & $0.4\pm0.9$    & $-7\pm19$  \\
HD64299   & 1997-09-01 & $0.34\pm0.05$  & $-5\pm17$  \\
HD283812  & 1997-09-28 & $1.32\pm0.04$* & $33\pm7$   \\
          & 1997-12-04 & $1.38\pm0.09$  & $29\pm5$   \\
HD331891  & 1997-09-01 & $0.36\pm0.05$  & $6\pm11$   \\
          & 1998-04-26 & $0.33\pm0.04$  & $5\pm13$   \\
          & 2002-09-09 & $0.26\pm0.04$  & $-2\pm11$  \\
          & 2003-06-08 & $0.26\pm0.03$  & $2\pm15$   \\ \hline
\enddata
\tablecomments{
Epochs marked with ``*'' are the results from the 1st quadrant of NIC2 only.
}
\end{deluxetable}

Adjusting the transmission co-efficients to null the measured polarization in a single unpolarized standard may provide a ``quick fix'' 
for some data, but it is by no means a concrete solution to the underlying problem. After all, it is possible there have been no 
changes in the transmission co-efficients. Ideally we would be able to characterize these effects by observing several polarized 
standards at the cardinal angles of the polarizers (this greatly reduces the degrees of freedom in the coefficient matrix). This would 
also allow us to see if there is an instrumental polarization that is dependent on the orientation and level of polarization in the object 
as well as the roll angle of {\it HST}. For example, it is possible that the sensitivity of the mirrors to polarized light is producing 
some reflective retardance. Unfortunately, there is, at present, insufficient calibration data available in the archive with which to 
fully check this possibility. There does exist data for two polarized objects, CHA-DC-F7 and HD283812, at multiple roll angles 
(see Table~\ref{tab:sample}), but this statistically insignificant sample inhibits us from drawing any conclusions about the nature
of an observationally dependent instrumental polarization. 

While previous polarimetric studies with NICMOS have not needed to be concerned with such characteristics due to the high levels of 
observed polarization \citep{umandm05}, as the capabilities of NICMOS are stretched by more and more ambitious programs so will the 
need to accurately measure low levels of polarization. In fact, a number of programs needing these levels of accuracy have already been 
executed. For example, \#10160 (The nuclear scattering geometry of Seyfert galaxies) and \#10410 (Anisotropy and obscuration 
in the near nuclear regions of powerful radio galaxies) will both require high accuracies as AGN generally display $p\approx1-5\%$. 
In addition, as the James Webb Space Telescope will not be flying with polarimetry optics, NICMOS will remain the only instrument 
capable of performing such high precision imaging polarimetry on faint objects.

The recent polarization calibration for the Advanced Camera for Surveys \citep{bir04,bandkp04,kpandb04,kpandb05} presents an example for 
the NICMOS calibration. In order for the aforementioned programs to be successful, new data on a number of both polarized standards, 
at various degrees and orientations of polarization, and unpolarized standards should be obtained. Pointings should be gridded across the 
detector in order to investigate field dependence. A number of exposures, in each polarizer, at different roll angles should also be 
employed in order to replicate the ground characterization. This will allow the derivation of element independent transmission 
co-efficients (from point sources) and will allow tests for higher order calibration effects. Such thorough observations may also shed 
some light as to the reason for the change in behavior from Cycles 7 and 11.

\section{Conclusions}\label{cons}

We have conducted polarimetric analyses on all (un)polarized standards available in the NICMOS data archive for the NIC2 camera. The 
principle aim was to determine the behavior of the instrument at low levels of polarization. We have thoroughly tested our routine for 
deriving aperture polarimetry of objects and found it to be robust. It has also demonstrated that observed small radius variation 
in the polarimetric curves of growth can be attributed to the effects of sub-pixel mis-alignments between polarizing elements and the 
point spread function. We have found no evidence for persistence.

Our findings also indicate that there is a measured polarization of $p\approx1.2\%$ in unpolarized targets which may indicate an 
intrinsic instrumental polarization and the need for further tweaking of the on-orbit transmission co-efficients. Assuming we are 
detecting an instrumental effect we have corrected data for polarized targets in the ($Q,U$)-plane and found the dispersion around 
previous (ground based) polarization estimates to have been reduced, but not totally removed. We have also derived averaged values of 
$t_k$ that null the unpolarized standard HD331891 across all apertures, and found $t_1$ and $t_2$ to be 0.8717 and 0.8341, respectively, 
when fixing $t_3=0.9667$. In addition, we have attempted to investigate the possibility of an observationally dependent instrumental 
polarization, but are inhibited from any conclusions by an insignificant sample.

While such levels of residual intrinsic polarization may not hamper studies of highly polarized targets, these levels will have a 
detrimental effect on studies attempting to measure $p<5\%$. It is clear that a more comprehensive calibration study of NICMOS is 
critical in order for a number of {\it HST} programs to be carried out successfully. The current post-NCS calibration archive (one 
polarized and one unpolarized standard) is insufficient for this to occur effectively. An array of polarized and unpolarized standards 
should be observed, at many roll angles and in all quadrants of the instrument, in order for the low level polarization characteristics 
of NICMOS to be properly and fully investigated, understood and removed.

\acknowledgments

We would like to thank the referee for his careful and thorough reading of this manuscript. His input has greatly improved the paper. 
Thanks are also extended to Glenn Schneider for his useful comments. Support for this study was provided by NASA through a grant from the 
Space Telescope Science Institute, which is operated by the Association of Universities for Research in Astronomy, Incorporated, under 
NASA contract NAS5-26555.

\end{document}